\documentclass[11pt,a4paper]{article}

\usepackage[a4paper,text={450pt,650pt},centering]{geometry}
\usepackage{hyperref}
\usepackage{mathrsfs}
\usepackage{amsmath,amssymb}
\usepackage{verbatim}
\usepackage{stmaryrd}
\usepackage{amsfonts,xcolor,ulem}

\newcommand{\alg}[1]{\mathfrak{#1}}

\newcommand{\el}{\nonumber\\}

\def\ads{{\rm AdS}_5\times {\rm S}^5}

\def\xp{{x^+}}
\def\xm{{x^-}}

\def\sl{\mathfrak{sl}}
\def\gl{\mathfrak{gl}}

\def\afQ{\widehat{\cal{Q}}}

\def\glone{\mathcal{U}_q(\widehat{\mathfrak{gl}}(1|1))}
\def\gltwo{\mathcal{U}_q(\widehat{\mathfrak{gl}}(2|2))}
\def\sltwo{\mathcal{U}_q(\widehat{\mathfrak{sl}}(2|2))}

\def\ads{{\rm AdS}_5\times {\rm S}^5}

\numberwithin{equation}{section}

\begin{document}


\begin{titlepage}

\begin{flushright}\small{DMUS-MP-11/02}\end{flushright}


\begin{centering}

\vspace{1in}

{\Large {\bf The Quantum Affine Origin of the AdS/CFT Secret Symmetry}}

\vspace{.5in}

{\large Marius de Leeuw${}^{a,1}$, Vidas Regelskis${}^{b,c,2}$ and Alessandro Torrielli${}^{d,3}$}\\
\vspace{.3 in}
${}^{a}${\textit{ETH Z\"urich, Institut f\"ur Theoretische Physik, \\Wolfgang-Pauli-Str.\ 27, CH-8093 Zurich, Switzerland}} \\
${}^{b}${\textit{Department of Mathematics, University of York,\\Heslington, York YO10 5DD, UK}}\\
${}^{c}${\textit{Institute of Theoretical Physics and Astronomy of Vilnius University,\\Go\v{s}tauto 12, Vilnius 01108, Lithuania}}\\
${}^{d}${\textit{Department of Mathematics, University of Surrey,\\Guildford, Surrey, GU2 7XH, UK}}\\

\footnotetext[1]{{\tt deleeuwm@phys.ethz.ch,}\quad ${}^{2}${\tt vr509@york.ac.uk,} ${}^{3}${\tt a.torrielli@surrey.ac.uk}\quad }
\vspace{.5in}

\vspace{.1in}

\end{centering}

\abstract{
We find a new quantum affine symmetry of the S-matrix of the one-dimensional Hubbard chain. We show that this symmetry originates from the quantum affine superalgebra $\gltwo$, and in the rational limit exactly reproduces the secret symmetry of the AdS/CFT worldsheet S-matrix.
}

\end{titlepage}

\tableofcontents


\section{Introduction}

Recent progress in exploring the quantum deformed one-dimensional Hubbard model \cite{BK} was inspired by the construction of the (doubly) deformed quantum affine algebra $\afQ$ \cite{BGM}. The algebra $\afQ$ may be viewed as the affine lift of the centrally extended superalgebra $\alg{sl}(2|2)$ that governs the worldsheet S-matrix of the AdS/CFT correspondence\footnote{For a review, see \cite{BetAl}. Note also that we will not specify a real form of the algebra throughout this paper, although it will always be possible to specialize to such a choice at any stage.} \cite{BDynamic,BAnalytic}. This relation implies that many of the properties of the worldsheet S-matrix can be reinterpreted in the light of their quantum affine lift. For example, the Yangian symmetry of the worldsheet S-matrix \cite{BYangian} is equivalent to the rational $q\to1$ limit of $\afQ$. In a similar way the bound state worldsheet S-matrices \cite{ALT_BStates,ALT_Rmat} can be obtained from the bound state S-matrix of the deformed Hubbard chain \cite{LMR}. 

One of the most peculiar features of the worldsheet S-matrix is the so-called secret symmetry, which acts as a helicity operator on the states and would normally extend the superalgebra $\alg{su}(2|2)$ to $\alg{gl}(2|2)$ \cite{MMT}. This symmetry was shown to be present only at the Yangian level, since the corresponding Lie algebra charge is not a symmetry of the worldsheet S-matrix. It is important to note that symmetries of a similar origin were found in quite a few related models. For instance, it revealed itself as a twisted secret symmetry of the boundary scattering K-matrices \cite{Regelskis} and it also appears as a so-called `bonus' Yangian symmetry for scattering amplitudes in $\mathcal{N}=4$ SYM \cite{BS,BM}. Thus, the secret symmetry should perhaps be regarded as an integral part of the symmetries of the model. The need for such an extension responds to a consistency issue of the underlying quantum group description of the integrable structure, according to a general prescription by Khoroshkin and Tolstoy \cite{KT}. According to this mathematical argument, in the case of superalgebras with a degenerate Cartan matrix, one may adopt the R-matrix of the smallest non-degenerate algebra containing the original one. The universal R-matrix found in such a way intertwines {\it a fortiori} the coproducts of the original algebra. This expectation leads to the natural question of whether a similar symmetry is also hidden in the S-matrices of the deformed Hubbard chain.

A first hint that this is the case is found in the classical limit of the corresponding integrable structure. For the undeformed theory the secret symmetry plays a crucial role in the classical limit, where it is needed to achieve factorization of the classical $r$-matrix in the form of a quantum-double \cite{Aleclass,AleSaneclass,Bclass,dLclass}. Similarly, the secret symmetry generator is also appearing in the universal expression of the $q$-deformed $r$-matrix \cite{BclassAff}. This seems to point towards an affirmative answer to the question of whether this additional symmetry is present for the full quantum $\afQ$-invariant S-matrix.

A natural place to start investigating is the so-called `conventional' affine limit of the algebra $\afQ$. This limit is obtained by sending one of the (complex) parameters, namely, the coupling $g$, to zero, followed by a suitable twist that removes the braiding factors inherited from the rational case \cite{GH,PST}. In this limit two of the three central charges of $\afQ$ vanish; thus, the algebra becomes isomorphic to the regular quantum affine universal enveloping superalgebra $\sltwo$. By adjoining the non-supertraceless Cartan generators $h_{4,0}$ and $h_{4,\pm1}$, one may extend $\sltwo$ to $\gltwo$. The representations of $\gltwo$ may be obtained using the general prescription for the superalgebras $\mathcal{U}_q(\widehat{\alg{gl}}(n|n))$, as presented in \cite{Zhang}. 

The R-matrix in this conventional limit is naturally found to respect the full $\gltwo$ symmetry. In other words, one automatically finds an extended symmetry, corresponding to the hypercharge operators $h_{4,i}$. However, turning the coupling back on, we see the appearance of the same phenomenon as in the rational case: The level one non-supertraceless generator is once again preserved (and, presumably, all higher levels), while the level zero is broken. We find two secret symmetries that we call $B_E$ and $B_F$ of the (bound state) S-matrices of deformed Hubbard chain. These symmetries are the natural (although non-trivial) analog of the Cartan generators $h_{4,\pm1}$ of $\gltwo$. In the rational $q\to1$ limit they exactly reproduce the secret symmetry of the worldsheet S-matrix \cite{MMT}.

The paper is organized as follows. In section 2 we review the Chevalley-Serre and Drinfeld's second realization of the quantum affine superalgebra $\glone$. We consider its fundamental representation and give the explicit realization of the corresponding R-matrix and the non-supertraceless charges $h_{2,0}$ and $h_{2,\pm1}$. This can be considered both as a warm-up exercise, and as a treatment relevant to a wealth of subsectors of the full algebra and corresponding R-matrix, later discussed in Section 4. In section 3 we review the superalgebra $\gltwo$ and its fundamental representation, and give the necessary background for building the secret symmetry of $\afQ$. In section 4, bearing on the construction presented in section 3, we build the Secret symmetry of the bound state S-matrices of $\afQ$ in both the conventional limit ($g\to0$) and the full deformed case. In the Conclusions, we present future directions and comment on the connection between the presence of the secret symmetry and the consistency of integrable sigma models based on superalgebras with vanishing Killing form.


\section{The quantum affine superalgebra \texorpdfstring{$\glone$}{Uq(gl(1|1))} }\label{Sec:GL11}

In this section we provide both the Chevalley-Serre realization and the so called Drinfeld's second realization \cite{Drin} of the quantum affine superalgebra $\glone$, in the conventions of \cite{Zhang} (see also \cite{Yama1,Yama2,Zhang2,Zhang3,BGZD,DGLZ}). We choose a complex number $q \neq 0$ and not a root of unity, and define
\begin{align}
[y]_q = \frac{q^y-q^{-y}}{q-q^{-1}} \,.
\end{align}
We will also set the central charge $c$ of the quantum affine algebra to zero for the rest of this section, and generically indicate with $[\;,\,]$ the graded commutator. 


\subsection{Chevalley-Serre realization}

In the Chevalley-Serre realization, $\glone$ is generated by fermionic Lie superalgebra positive (respectively, negative) roots $\xi^+_1$ (respectively, $\xi^-_1$), Cartan generators $h_1,h_2$, with $h_2$ the non-supertraceless element completing the Lie superalgebra $\sl(1|1)$ to $\gl(1|1)$, and the affine fermionic positive (respectively, negative) roots $\xi^+_0$ (respectively, $\xi^-_0$) and corresponding Cartan generator $h_0\,$. 

The generalized symmetric Cartan matrix is given by
\begin{align}
(a_{ij})_{0 \leqq i,j \leqq 2}=
\left(\begin{array}{cccc}
0&0&-2\\
0&0&2\\
-2&2&0
\end{array}\right) \,.
\end{align}
Notice that this matrix is degenerate, but the Lie superalgebra block $1 \leqq i,j \leqq 2$ is not.

The defining relations are as follows, for $0 \leqq i,j \leqq 2$ (root generators corresponding to the Cartan generator $h_2$ are absent):
\begin{align}
[h_i,h_j] &= 0\,, & [h_i,\xi^\pm_j] &= \pm a_{ij}x^\pm_j\,, & \{\xi^+_i,\xi^-_j\} &= \delta_{ij}\frac{q^{h_i}-q^{-h_i}}{q-q^{-1}}\,,
\end{align}
supplemented by a suitable set of Serre relations. We refer to \cite{Zhang} for the explicit form of the Serre relations, as we will instead spell out the complete set of relations in Drinfeld's second realization, see (\ref{DII}).

One can define a Hopf algebra structure with the following coproduct, counit and antipode:
\begin{align}
\label{cop1}
\Delta (h_i) &= h_i \otimes 1 + 1 \otimes h_i\,,                    & S(h_i) &= - h_i\,, \el
\Delta (\xi^+_i) &= \xi^+_i \otimes 1 + q^{h_i} \otimes \xi^+_i\,,  & S(\xi^\pm_i) &= - q^{\mp h_i} \xi^\pm_i\,, \el
\Delta (\xi^-_i) &= \xi^-_i \otimes q^{-h_i} + 1 \otimes \xi^-_i\,, & \epsilon (h_i)  &= \epsilon(\xi^\pm_i) =0\,.
\end{align}


\subsection{Drinfeld's second realization}

The same algebra is also generated by an infinite set of Drinfeld's generators, which in some sense make explicit the infinite set of `levels' of the quantum affine algebra obtained, in the Chevalley-Serre realization, by subsequent commutation with the affine roots $\xi^\pm_0$. These generators are  
\begin{align}
\xi_{1,m}^\pm\,, ~ h_{i,n}\,, \qquad\text{with}\qquad i=1,2\,, \quad m,n \in {\mathbb Z}\,.
\end{align}
The defining relations are as follows:
\begin{align}
{[h_{i,m} \,,\, h_{j,n}]} &= 0\,, & \{\xi_{1,n}^+ \,,\, \xi_{1,m}^-\} &= \frac{1}{q-q^{-1}} \left( \psi_{1,n+m}^+ -\psi_{1,n+m}^- \right), \quad \el
{[h_{i,0} \,,\, \xi_{1,m}^{\pm}]} &= \pm a_{i1} \, \xi_{1,m}^\pm\,, & \{\xi^\pm_{1,n} \,,\, \xi^\pm_{1,m}\} &= 0\,, \el
{[h_{i,n} \,,\, \xi_{1,m}^\pm]} &= \pm \frac{[a_{i1} n]_q}{n} \, \xi_{1,n+m}^\pm\,, & \text{for} \quad n \neq 0\,.  \label{DII}
\end{align}
We have used the definition
\begin{align}
\label{psii}
\psi_1^\pm(z) &= q^{\pm h_{1,0}}
\exp\left(\pm
(q-q^{-1})\sum_{m>0}h_{1,\pm m}z^{\mp m}\right) = \sum_{n \in \mathbb Z } \psi_{1,n}^\pm z^{-n}\,.
\end{align}
The above expression (\ref{psii}) should be understood as defining a generating function for the individual $\psi_{1,n}^\pm$'s, which in turn can be obtained by Laurent expanding both sides of the equation and matching the powers of the parameter $z$.
 
We call `level' the index $n$ of Drinfeld's generators. One typically introduces a `derivation' operator $d$ that counts the level, in the following way:
\begin{align}
\label{deriv}
[d,\tau_n] = n \, \tau_n \,,
\end{align}
for any generator $\tau_n$ at level $n$.

The map between the Chevalley-Serre and Drinfeld's second realization is given by the following assignment, for $i=1,2$:
\begin{align}
\label{D1map}
h_i & = h_{i,0} \,,   \qquad\qquad \xi^\pm_1 = \xi^\pm_{1,0} \,, \nonumber\\
h_0 & = - h_{1,0} \,, \qquad\quad\, \xi^\pm_0 = \pm \xi^\mp_{1,\pm 1} \, q^{\mp h_{1,0}} \,,
\end{align}
where we have used the fact that $a_{11}=0$. As one can see, the positive (respectively, negative) affine root in the Chevalley-Serre realization generates the positive (respectively, negative) tower of levels in Drinfeld's second realization. 

The coalgebra structure in Drinfeld's second realization satisfies the following triangular decomposition, for $n \in {\mathbb Z}, n \neq 0$ (for $n=0$ the coproduct can be obtained directly from (\ref{D1map}), (\ref{cop1})):
\begin{align}
\label{triang}
\Delta(h_{1,n}) & = h_{1,n} \otimes 1 + 1 \otimes h_{1,n} \, \, mod  \, \, N_- \otimes N_+, \el
\Delta(\xi^+_{1,n}) & = \xi^+_{1,n} \otimes 1 + q^{sign(n) h_{1,0}} \otimes \xi^+_{1,n} \el
 & \quad + \, \sum_{k=\frac{1}{2}(1-sign(n))}^{|n|-1} \psi_{1,sign(n)(|n|-k)}^{sign(n)} \otimes \xi^+_{1,sign(n) k} \, \, mod \, \, N_- \otimes N_+^2,\el
\Delta(\xi^-_{1,n}) & = \xi^-_{1,n} \otimes q^{sign(n) h_{1,0}} + 1 \otimes \xi^-_{1,n} \el
 & \quad + \, \sum_{k=\frac{1}{2}(1+sign(n))}^{|n|-1} \xi^-_{1,sign(n) k} \otimes \psi_{1,sign(n)(|n|-k)}^{sign(n)} \, \, mod \, \, N_-^2 \otimes N_+,
\end{align}
with $N_\pm$ (respectively, $N_\pm^2$) the left ideals generated by $\xi^\pm_{1,m}$ (respectively, $\xi^\pm_{1,m} \xi^\pm_{1,m'}$), with $m,m' \in {\mathbb Z}$.

The coproduct for the generators $h_{2,n}$ is obtained by imposing that $\Delta$ is an algebra homomorphism, namely, that it respects the defining relations (\ref{DII}). Making use of (\ref{triang}), we obtain for instance
\begin{align}
\Delta h_{2,+1} & = h_{2,+1} \otimes 1 + 1 \otimes h_{2,+1} + (q^{-2}-q^2) \xi^{-}_{1,+1} \otimes \xi^{+}_{1,0} \,, \el 
\Delta h_{2,-1} & = h_{2,-1} \otimes 1 + 1 \otimes h_{2,-1} - (q^{-2}-q^2) \xi^{-}_{1,0} \otimes \xi^{+}_{1,-1} \,. \label{h11_DII}
\end{align}
%


\subsection{Fundamental representation}

We provide here what we will call the `fundamental' representation in Drinfeld's second realization, as obtained from \cite{Zhang} by specializing to a particular case. To obtain the corresponding representation in the Chevalley-Serre realization, one can make use of Drinfeld's map (\ref{D1map}). For $v_1$ and $v_2$ a bosonic and fermionic state, respectively, $\eta_{ij}$ the matrix with $1$ in position $(i,j)$ and zero elsewhere, and $z$ a spectral parameter counting the level, we have for instance
\begin{align}
\label{fond}
\xi^+_{1,0} &= \eta_{12} \,,  & \xi^-_{1,0} &= \eta_{21} \,, \hspace{1in} h_{1,0} = \eta_{11} + \eta_{22} \,, \el
h_{2,0} &= \eta_{11} - \eta_{22} \,, & \qquad h_{2,\pm1} &= \tfrac{1}{2}(z\,q)^{\pm1} [2]_q \, (\eta_{11} - \eta_{22}) \,, \hspace{1.5in} \el
\xi^+_{1,\pm1} &= (z \, q)^{\pm1} \, \eta_{12} \,,  & \xi^-_{1,\pm1} &= (z \, q)^{\pm1} \, \eta_{21} \,.
\end{align} 

The derivation (\ref{deriv}) in this (so-called `evaluation') representation is given by $d = z \tfrac{d}{dz}$. The R-matrix satisfying the invariance condition
\begin{align}
\label{invariante}
\Delta^{op} (\tau) \, R \, = R \, \Delta (\tau) \,,
\end{align}
with $\Delta^{op} (\tau)$ defined as $\Delta (\tau)$ followed by a graded permutation, and $\tau$ any generator of the algebra, is given (up to an overall factor) by (see also \cite{CaiEtAl})
\begin{align}
\label{RZhang}
R &= \eta_{11} \otimes \eta_{11} + \frac{\frac{z}{w} -1}{q \frac{z}{w} - q^{-1}}  (\eta_{11} \otimes \eta_{22} + \eta_{22} \otimes \eta_{11})\nonumber\\
 & \quad + \, \frac{\frac{z}{w}(q - q^{-1})}{q \frac{z}{w} - q^{-1}}  (\eta_{21} \otimes \eta_{12} - \frac{w}z{} \eta_{12} \otimes \eta_{21}) + \frac{q^{-1} \frac{z}{w} - q}{q \frac{z}{w} - q^{-1}}  \eta_{22} \otimes \eta_{22} \,,
\end{align}
where $z,w$ are the spectral parameters corresponding to the first and second copy of the algebra respectively. \\

Finally, we want to translate the expressions \eqref{h11_DII} into the Chevalley-Serre basis, as this shall be important to us later on. This can be done with the help of \eqref{D1map}. However, the charges $h_{2,\pm1}$ have no canonical image under Drinfeld's map. For this reason, let us introduce new charges
\begin{align}
  \mathcal{B_\pm} = \frac{\,(z \, q)^{\pm1}}{q^{-1}-q} \, h_{2,0}. \label{B_GL11}
\end{align}
In the Chevalley-Serre basis, \eqref{h11_DII} then reads as
\begin{align}
\Delta \mathcal{B}_+ & = \mathcal{B}_+ \otimes 1 + 1 \otimes \mathcal{B}_+ + 2 \, \xi^{+}_{0} k_1 \otimes \xi^{+}_{1}, \el 
\Delta \mathcal{B}_- & = \mathcal{B}_- \otimes 1 + 1 \otimes \mathcal{B}_- + 2 \, \xi^{-}_{1} \otimes k_1^{-1} \xi^{-}_{0}. \label{DB11}
\end{align}


\section{The quantum affine superalgebra \texorpdfstring{$\gltwo$}{Uq(gl(2|2))} }\label{Sec:GL22}

We will now specialize the presentation of \cite{Zhang} to the case of $\gltwo$. While the previous section is strictly related to certain subsectors of the $q$-deformed AdS/CFT algebra (which we will treat in the second part of the paper), this section is related to the full algebra and corresponding R-matrix. We will directly focus on Drinfeld's second realization for simplicity, referring to \cite{Zhang} for further details (see also \cite{YamaMulti}).


\subsection{Drinfeld's second realization}

The algebra $\gltwo$ (for an all-fermionic Dynkin diagram) is generated by an infinite set of Drinfeld's generators  
\begin{align}
\xi_{i,m}^\pm \,, ~h_{j,n} \,, \qquad\text{with}\qquad i=1,2,3, ~~~ j=1,2,3,4, ~~~ m,n \in {\mathbb Z}.
\end{align}
The defining relations are as follows:
\begin{align}
\label{DII22}
& {[h_{j,m} \,,\, h_{j',n}]} = 0 \,, \el
& {[h_{j,0} \,,\, \xi_{i,m}^\pm]} = \pm a_{ji} \, \xi_{i,m}^\pm \,, \el
& {[h_{j,n} \,,\, \xi_{i,m}^\pm]} = \pm \frac{[a_{ji} \, n]_q}{n} \, \xi_{i,n+m}^\pm \,,  \quad n \neq 0 \,, \el
& \{\xi_{i,n}^+ \,,\, \xi_{i',m}^-\}
=\frac{\delta_{i,i'}}{q-q^{-1}}
\left(
\psi_{i,n+m}^+ -
\psi_{i,n+m}^-
\right),
\end{align}
combined with a suitable set of Serre relations \cite{Zhang} which read 
\begin{align}
\label{Ser}
&\{\xi_{i,m}^\pm \,,\, \xi_{i',n}^\pm\} = 0, \quad \text{if} \quad a_{ii'} = 0 \,, \el
&\{\xi_{i,m+1}^\pm \,,\, \xi_{i',n}^\pm\}_{q^{\pm a_{ii'}}}=\{\xi_{i',n+1}^\pm, \xi_{i,m}^\pm\}_{q^{\pm a_{ii'}}} \,,\\
&[\{\xi_{2,m}^\pm \,,\, \xi_{1,n}^\pm\}_q,\{\xi_{2,p}^\pm \,,\, \xi_{3,r}^\pm\}_{q^{-1}}]=[\{\xi_{2,p}^\pm \,,\, \xi_{1,n}^\pm\}_q,\{\xi_{2,m}^\pm \,,\, \xi_{3,r}^\pm\}_{q^{-1}}] \,. \nonumber
\end{align}
The symmetric Cartan matrix reads
\begin{align}
(a_{ij})_{1 \leqq i,j \leqq 4}=
\left(\begin{array}{cccc}
0&1&0&2\\
1&0&-1&-2\\
0&-1&0&2\\
2&-2&2&0
\end{array}\right) \,.
\end{align}
We have once again used the definition
\begin{align}
\psi_i^\pm(z) &= q^{\pm h_{i,0}}
\exp\left(\pm
(q-q^{-1})\sum_{m>0}h_{i,\pm m}z^{\mp m}\right) = \sum_{n \in \mathbb Z } \psi_{i,n}^\pm z^{-n} \,.
\end{align}
The `derivation' operator $d$ counting the level is once again introduced in the following way:
\begin{align}
\label{deriv22}
[d,\tau_n] = n \, \tau_n,
\end{align}
for any generator $\tau_n$ at level $n$.

Let us comment on the Serre relations (\ref{Ser}). The first line expresses the fermionic nature of the simple generators, while the second one ensures that a good filtration is preserved. This means that one is free to combine levels in different ways to obtain one and the same `sum' level as a result. The third line, taken at level $0$ (namely, for $m=n=p=r=0$), tells us that there are only three non-simple roots, two obtained as $\{ \xi_{2,0}^\pm, \xi_{1,0}^\pm\}_q$ and $\{ \xi_{2,0}^\pm, \xi_{3,0}^\pm\}_{q^{-1}}$, and one obtained by commuting, for instance, the very first of these non-simple roots with $\xi_{3,0}^\pm$. In fact, the third Serre relation implies that commuting the two non-simple roots with each other returns zero, which truncates any further generation of roots.  

The coproduct has the natural structure (we define $sign(0) \equiv +1$)
\begin{align}
\label{triang22}
\Delta(h_{i,n}) & = h_{i,n} \otimes 1 + 1 \otimes h_{i,n} \, \, mod  \, \, N_- \otimes N_+ \,, \el
\Delta(\xi^+_{i,n}) & = \xi^+_{i,n} \otimes 1 + q^{sign(n) h_{i,0}} \otimes \xi^+_{i,n} \el
 & \quad + \, \sum_{k=\frac{1}{2}(1-sign(n))}^{|n|-1} \psi_{i,sign(n)(|n|-k)}^{sign(n)} \otimes \xi^+_{i,sign(n) k} \, \, mod \, \, N_- \otimes N_+^2 \,,\el
\Delta(\xi^-_{i,n}) & = \xi^-_{i,n} \otimes q^{sign(n) h_{i,0}} + 1 \otimes \xi^-_{i,n} \el
 & \quad + \, \sum_{k=\frac{1}{2}(1+sign(n))}^{|n|-1} \xi^-_{i,sign(n) k} \otimes \psi_{i,sign(n)(|n|-k)}^{sign(n)} \, \, mod \, \, N_-^2 \otimes N_+ \,,
\end{align} 
with $N_\pm$ (respectively, $N_\pm^2$) the left ideals generated by $\xi^\pm_{i,m}$ (respectively, $\xi^\pm_{i,m} \xi^\pm_{i,m'}$), with $m,m' \in {\mathbb Z}$ and $i=1,2,3$.

The coproduct for the generators $h_{4,n}$ is obtained by imposing that $\Delta$ respects the defining relations (\ref{DII22}). With respect to the case of $\glone$, the `tail' of the coproduct ({\it i.e.}, the quadratic part that comes after the trivial comultiplication rule for the generator itself) now contains non-simple roots (which before where simply absent). By carefully taking into account (\ref{triang22}), we find
\begin{align}
\Delta(h_{4,+1}) &= h_{4,+1} \otimes 1 + 1 \otimes h_{4,+1} + (q^{-1} - q) \sum_{i=1}^3 \, [a_{4i}]_q \, \xi^-_{i,+1} \otimes \xi^+_{i,0} \, + \, \mbox{\footnotesize{non-simple roots}} \,,\el
\Delta(h_{4,-1}) &= h_{4,-1} \otimes 1 + 1 \otimes h_{4,-1} - (q^{-1} - q) \sum_{i=1}^3 \, [a_{4i}]_q \, \xi^-_{i,0} \otimes \xi^+_{i,-1} \, + \, \mbox{\footnotesize{non-simple roots}} \,. \label{h41_GL22}
\end{align}
We will specify the non-simple part of the tail of the coproduct in the fundamental representation in the following section.


\subsection{Fundamental representation}

The `fundamental' representation in Drinfeld's second realization can be obtained from \cite{Zhang} in a particular case. For {\rm $v_1,v_2$ and $v_3,v_4$ two bosonic and two fermionic states, respectively}, $\eta_{ij}$ the matrix with $1$ in position $(i,j)$ and zero elsewhere, and $z$ a spectral parameter counting the level, we have this time
\begin{align}
& \begin{aligned}
\xi^+_{1,0} &= \eta_{13} \,, & \xi^+_{2,0} &= \eta_{32} \,,   & \xi^+_{3,0} &= \eta_{24} \,, \el
\xi^-_{1,0} &= \eta_{31} \,, & \xi^-_{2,0} &= - \eta_{23} \,, & \xi^-_{3,0} &= \eta_{42} \,, \el
h_{1,0} &= (\eta_{11} + \eta_{33}), & h_{2,0} &= - (\eta_{33} + \eta_{22}), & h_{3,0} &= (\eta_{22} + \eta_{44})\el
h_{4,0} &= \sum_{k=1}^4 (-)^{[k]} \eta_{kk} \,, \\
\xi^+_{1,\pm1} &= (z \, q)^{\pm1} \, \eta_{13} \,, & \xi^+_{2,\pm1} &= z^{\pm1} \, \eta_{32} \,,   & \xi^+_{3,\pm1} &= (z \, q)^{\pm1} \, \eta_{24} \,, \el
\xi^-_{1,\pm1} &= (z \, q)^{\pm1} \, \eta_{31} \,, & \xi^-_{2,\pm1} &= - z^{\pm1} \, \eta_{23} \,, & \xi^-_{3,\pm1} &= (z \, q)^{\pm1} \, \eta_{42} \,, \el 
h_{1,\pm1} &= z^{\pm1} (\eta_{11} + \eta_{33}) \,, & h_{2,\pm1} &= -(z\,q)^{\pm1} (\eta_{22} + \eta_{33}) \,, & h_{3,\pm1} &= z^{\pm1} (\eta_{22} + \eta_{44}) \,,  \el
\end{aligned} \\
& h_{4,\pm1} = z^{\pm1} \, [2]_q \, \Big( y^{\pm} \eta_{11}  + (y^{\pm}+1-q^{\pm1}) \eta_{22}  + (y^{\pm}-q^{\pm1}) \eta_{33}  + (y^{\pm} + 1 - 2 q^{\pm1}) \eta_{44} \Big) \label{fund22} \,, 
\end{align}
with $[k]$ the grading of the state $v_k$. The derivation (\ref{deriv22}) in the evaluation representation (\ref{fund22}) is given by $d = z\,\tfrac{d}{dz}$. 
The algebra $\gl(n|n)$ is non-semisimple ($\sl(n|n)$ being a non-trivial ideal strictly contained in it). Hence, one can always add a constant times the identity to the non-supertraceless generator who lives outside the ideal (and, therefore, never appears on the right-hand-side of any commutation relations). The generator $h_{4,1}$ of the quantum-affine version also does not appear on the r.h.s. of any commutation relations, and one can use the freedom we just mentioned to redefine this generator by adding a multiple of the identity. This is reflected in the choice of $y^\pm$ (which we tacitly fixed to a convenient value in the previous section). The term multiplying $y^\pm$ is a multiple of the identity matrix, and its coproduct is trivial hence it drops out of the defining relation for the R-matrix (\ref{invariante}). 

Let us spell out the coproduct (\ref{h41_GL22}) in this representation ($z$ and $w$ once again refer to the first and, respectively, the second factor in the tensor product):
\begin{align}
\label{h4122compl}
\Delta h_{4,+1} & = h_{4,+1} \otimes 1 + 1 \otimes h_{4,+1} + (q^{-2}-q^2) \, z \, \Big( q \eta_{31} \otimes \eta_{13} + (q-1) \eta_{21} \otimes \eta_{12} \el
 & \quad + (2q-1) \eta_{41} \otimes \eta_{14} +  \eta_{23} \otimes \eta_{32} + (1-q) \eta_{43} \otimes \eta_{34} + q \eta_{42} \otimes \eta_{24} \Big). \el
\Delta h_{4,-1} & = h_{4,-1} \otimes 1 + 1 \otimes h_{4,-1} - (q^{-2}-q^2) \, w^{-1} \, \Big( q^{-1} \eta_{31} \otimes \eta_{13} + (q^{-1}-1) \eta_{21} \otimes \eta_{12} \el
 & \quad + (2q^{-1}-1) \eta_{41} \otimes \eta_{14} +  \eta_{23} \otimes \eta_{32} + (1-q^{-1}) \eta_{43} \otimes \eta_{34} + q^{-1} \eta_{42} \otimes \eta_{24} \Big).
\end{align}
Notice that the bosonic part of the tail is higher order in the $q \to 1$ limit, and therefore it disappears in the Yangian limit. The parameter $y$ does not appear in the coefficients of the tail, according to the above discussion. We can once again fix the constant $y$ to a convenient value, for instance 
\begin{align}
y^{\pm} = q^{\pm1} - \tfrac{1}{2} \,, 
\end{align}
which produces the following representation:
\begin{align}
h_{4,\pm1} = z^{\pm1} \, [2]_q \, \Big( (q^{\pm1}-\tfrac{1}{2}) \eta_{11} + \tfrac{1}{2} \eta_{22} - \tfrac{1}{2} \eta_{33} - (q^{\pm1} -\tfrac{1}{2}) \eta_{44} \Big) \,,
\end{align}
The R-matrix satisfying the invariance condition (\ref{invariante}) is given (up to an overall factor) by (see also \cite{Gade})
\begin{align}
R & = \eta_{11}\otimes\eta_{11} + \eta_{22}\otimes\eta_{22} + \frac{q^{2}-\frac{z}{w}}{1-q^{2}\frac{z}{w}} \left(\eta_{33}\otimes\eta_{33}+\eta_{44}\otimes\eta_{44}\right) \el
 & \quad + \frac{q\left(1-\frac{z}{w}\right)}{1-q^{2}\frac{z}{w}} \sum_{i\neq j}\eta_{ii}\otimes\eta_{jj} - \frac{q^{2}-1}{q^{2}\frac{z}{w}-1} \left(\sum_{(i,j)\in A}\eta_{ij}\otimes\eta_{ji}-\eta_{12}\otimes\eta_{21}-\eta_{32}\otimes\eta_{23} \right) \el
 & \quad + \frac{q^{2}-1}{q^{2}-\frac{w}{z}}\left(\sum_{(i,j)\in B}\eta_{ij}\otimes\eta_{ji}-\eta_{23}\otimes\eta_{32}-\eta_{43}\otimes\eta_{34}\right), \label{R_GL22}
\end{align}
As a consistency check, one can notice that in the scaling limit $q = e^h$ and $z/w = e^{2\, \delta u\, h}$ 
with $h\to0$, the above R-matrix reduces to the Yangian R-matrix 
\begin{align}
R_Y =  \frac{\delta u}{\delta u+1} \left(1 + \frac{P}{\delta u}\right) ,
\end{align}
with $P$ being the graded permutation operator $P = \sum_{i,j=1}^4 (-)^j \,\eta_{ij} \otimes \eta_{ji}$ . \\

One can show that the combination
\begin{align}
\mathcal{B}_{\pm} = \frac{q^{\pm1}}{q^{-1}-q} \bigg(\frac{2}{q^{\pm1} [2]_q} h_{4,\pm1} \, + \, (q^{\mp1}-1) (h_{1,\pm1} - h_{3,\pm1})\bigg) \,, 
\end{align}
is such that, in the fundamental representation (\ref{fund22}), one obtains an analog of \eqref{B_GL11},
\begin{align} 
\label{B_GL22}
\mathcal{B}_{\pm} =  \frac{(z\,q)^{\pm1}}{q^{-1}-q} \, \sum_{i=1}^4 \, (-)^{[i]} \, \eta_{ii}\,.
\end{align}
Then, using \eqref{h4122compl} and 
\begin{align}
\Delta h_{1,+1} & = h_{1,+1} \otimes 1 + 1 \otimes h_{1,+1} + (q^{-1}-q) \, z \, (\eta\otimes\eta)_h \,, \el
\Delta h_{1,-1} & = h_{1,-1} \otimes 1 + 1 \otimes h_{1,-1} - (q^{-1}-q) \, w^{-1} \, (\eta\otimes\eta)_h \,, \el
\Delta h_{3,+1} & = h_{3,+1} \otimes 1 + 1 \otimes h_{3,+1} - (q^{-1}-q) \, z \, (\eta\otimes\eta)_h \,, \el
\Delta h_{3,-1} & = h_{3,-1} \otimes 1 + 1 \otimes h_{3,-1} + (q^{-1}-q) \, w^{-1} \, (\eta\otimes\eta)_h \,,
\end{align}
where
\begin{align}
(\eta\otimes\eta)_h = \eta_{21} \otimes \eta_{12} - \eta_{23} \otimes \eta_{32}  + \eta_{41} \otimes \eta_{14} + \eta_{43} \otimes \eta_{34} \,,
\end{align}
we find
\begin{align}
\label{Beta}
\Delta \mathcal{B}_{+} & = \mathcal{B}_{+} \otimes 1 + 1 \otimes \mathcal{B}_{+} + 2 \, z\,q \, (\eta_{31} \otimes \eta_{13} \, + \eta_{23} \otimes \eta_{32} \, + \eta_{41} \otimes \eta_{14} \, + \eta_{42} \otimes \eta_{24}) \,, \el
\Delta \mathcal{B}_{-} & = \mathcal{B}_{-} \otimes 1 + 1 \otimes \mathcal{B}_{-} + 2 \, (w\,q)^{-1} \, (\eta_{31} \otimes \eta_{13} \, + \eta_{23} \otimes \eta_{32} \, + \eta_{41} \otimes \eta_{14} \, + \eta_{42} \otimes \eta_{24}) \,.
\end{align}

As in the previous section, we translate these expressions into the Chevalley-Serre basis. The map between the Chevalley-Serre and Drinfeld's second realization, in the fundamental representation which is relevant to the present discussion, is given by the following assignment:
\begin{align}
\label{Dmap}
h_i &= h_{i,0}\,,   & \xi^\pm_i &= \xi^\pm_{i,0} \,, \nonumber\\
h_0 &= - h_{1,0}-h_{2,0}-h_{3,0}\,, \qquad\quad\, & \xi^\pm_0 &= \pm (q\,z)^{\pm1} [[\xi^\mp_{1,0},\xi^\mp_{2,0}],\xi^\mp_{3,0}] \, q^{\mp (h_{1,0}+h_{2,0}+h_{3,0})} \,.
\end{align}
Thus with the help of \eqref{fund22} we find
\begin{align}
\Delta \mathcal{B}_{+} & = \mathcal{B}_{+} \otimes 1 + 1 \otimes \mathcal{B}_{+} + 2\, \big( \xi_{0}^{+}k_{123}\otimes \xi_{123}^{+} + \xi_{012}^{+}k_{3}\otimes \xi_{3}^{+} \el & \hspace{2in} - q^{2}\xi_{013}^{+}k_{2}\otimes \xi_{2}^{+}+\xi_{230}^{+}k_{1}\otimes \xi_{1}^{+} \big) \,, \el
\Delta \mathcal{B}_{-} & = \mathcal{B}_{-} \otimes1+1\otimes \mathcal{B}_{-} + 2\, \big(\xi_{123}^{-}\otimes k_{123}^{-1}\xi_{0}^{-} + \xi_{3}^{-}\otimes k_{3}^{-1}\xi_{012}^{-} 
 \el & \hspace{2in} - q^{-2}\xi_{2}^{-}\otimes k_{2}^{-1}\xi_{013}^{-}+\xi_{1}^{-}\otimes k_{1}^{-1}\xi_{230}^{-} \big) \,, \label{DB_GL22}
\end{align}
where we have used the short-hand notation $k_{ijk} = k_i k_j k_k$ and $\xi_{ijk} = [[\xi_i,\xi_j],\xi_k]$. One can observe that these expressions can formally be written as
\begin{align}
\Delta \mathcal{B}_{+} & = \mathcal{B}_{+} \otimes 1 + 1 \otimes \mathcal{B}_{+} + 2 \sum_{\alpha \in \Phi_0} \, c_\alpha \, \xi_{\delta-\alpha} k_\alpha \otimes \xi_{\alpha} \,, \el
\Delta \mathcal{B}_{-} & = \mathcal{B}_{-} \otimes1+1\otimes \mathcal{B}_{-} + 2 \sum_{\alpha \in \Phi_0} \, c_\alpha \, \xi_{-\alpha} \otimes k_\alpha^{-1} \xi_{\alpha-\delta} \,, \label{DB_gen}
\end{align}
where $\Phi_0$ is the set of all positive non-affine roots, $\delta$ is the affine root and $c_\alpha$'s are complex parameters. \\

Let us make a final remark concerning the symmetry we have just obtained. We derived the coproduct (\ref{Beta}) starting from an all-fermionic Dynkin diagram, and the pattern of simple and non-simple roots which appear in the tail of the coproduct respects the original choice of Dynkin diagram. For later purposes, it will turn out to be convenient to work with a so-called {\it distinguished} Dynkin diagram. This is associated to a basis with only one fermionic root. The assignment of simple roots will be different and this will reflect on the non-simple roots appearing in the tail. In order to be able to match with the expressions we will later find, it is useful to perform a twist of the coalgebra structure (and of the corresponding R-matrix) in the spirit of \cite{KTTwisting} (see also \cite{Zhang2}), where it is explained that such twists may involve factors of the universal R-matrix itself. One can check that the following transformation
\begin{align}
\Psi &= \text{Id} - (q-q^{-1})(\eta_{23} \otimes \eta_{32} + \eta_{32} \otimes \eta_{23}) - \frac{w}{z} \, \eta_{22} \otimes \eta_{33} - \frac{z}{w} \, \eta_{33} \otimes \eta_{22}  \el
 & \qquad\; - \eta_{34} \otimes \eta_{43} - \eta_{43} \otimes \eta_{34} - \eta_{33} \otimes \eta_{44} - \eta_{44} \otimes \eta_{33} \,, \label{Psi}
\end{align}
is such that
\begin{align}
\Delta' = \Psi \, \Delta \, \Psi^{-1} \qquad \text{and} \qquad R' &= \Psi^{op} \, R \, \Psi^{-1}.
\end{align}
gives
\begin{align}
\label{Beta'}
\Delta'(\mathcal{B}_{+}) &= \mathcal{B}_{+} \otimes 1 + 1 \otimes \mathcal{B}_{+} + 2 \, z \, q \, (\eta_{31} \otimes \eta_{13} \, + \eta_{32} \otimes \eta_{23} \, + \eta_{41} \otimes \eta_{14} \, + \eta_{42} \otimes \eta_{24}) \,,\el
\Delta'(\mathcal{B}_{-}) &= \mathcal{B}_{-} \otimes 1 + 1 \otimes \mathcal{B}_{-} + \frac{2}{w\,q} \, (\eta_{31} \otimes \eta_{13} \, + \eta_{32} \otimes \eta_{23} \, + \eta_{41} \otimes \eta_{14} \, + \eta_{42} \otimes \eta_{24}) \,,
\end{align}
which is an analog of \eqref{Beta} for the case of the distinguished Dynkin diagram. The inverse of \eqref{Psi} can be explicitly calculated, and it reads
\begin{align}
\Psi^{-1} = \text{Id} + \tau_1 \, \eta_{22} \otimes \eta_{33} + \tau_2 \, \eta_{33} \otimes \eta_{22} + \tau_3 \, \eta_{23} \otimes \eta_{32} + \tau_4 \, \eta_{32} \otimes \eta_{23} \,,
\end{align}
with
\begin{align}
\tau_1 &= - \big( (1 - w/z) + (q^{-1}-q)^2 \big) \,\omega^{-1} \,, \qquad  \tau_3 = \tau_4 = (q - q^{-1})\, \omega^{-1} \,, \el
\tau_2 &= - \big( (1 - z/w) + (q^{-1}-q)^2 \big) \,\omega^{-1} \,,
\end{align}
and
\begin{align}
\omega = (1 - z/w)(1 - w/z) + (q^{-1}-q)^2 \,.
\end{align}

The non-supertraceless generator we have been focusing our attention on is what will be promoted to the secret symmetry of the full $q$-deformed AdS/CFT model in the next section. While, in the conventional case we have just been treating, this generator literally extends the superalgebra $\alg{su}(2|2)$ to $\alg{gl}(2|2)$, it will instead only appear at the first quantum-affine level in the subsequent treatment, in parallel to the rational case. The need for such an extension is however the same as in the conventional situation. Its presence corresponds to a consistency issue of the underlying quantum group description of the integrable structure, according to the prescription of Khoroshkin and Tolstoy \cite{KT}. In their analysis, an additional Cartan generator is needed to invert the otherwise degenerate Cartan matrix. In turn, the invertibility of the Cartan matrix allows one to write down the universal R-matrix, which appears to be in exponential form with precisely the inverse Cartan matrix appearing at the exponent (see also \cite{AF}).


\section{Deformed quantum affine algebra \texorpdfstring{$\afQ$}{Q}}\label{Sec:Q}

Having explored the fundamental representations of the algebras $\,\glone$ and $\,\gltwo$, we are now ready to turn to the quantum affine algebra $\afQ$ constructed in \cite{BGM}. We start by reviewing its bound state representations, put forward in \cite{LMR}. Then, bearing on the construction presented in the previous sections, we build the secret symmetry of the representations of $\afQ$ considered in \cite{LMR}. Finally we show that this new symmetry is a quantum analog of the secret symmetry discovered in \cite{MMT}.


\subsection{Chevalley-Serre realization}

The algebra $\widehat{\mathcal{Q}}$ of the quantum deformed one-dimensional Hubbard chain is a double deformation of the centrally extended affine superalgebra $\widehat{\mathfrak{sl}}(2|2)$ whose Dynkin diagram has two bosonic (1, 3) and two fermionic (2, 4) roots \cite{BGM}. It is generated by four sets of Chevalley-Serre generators $K_i\equiv q^{H_i}$, $E_i$, $F_i$ ($i=1,\,2,\,3,\,4$) and two sets of central elements $U_{k}$ and $V_{k}$ ($k=2,\,4$), with $U_{k}$ being responsible for the so-called `braiding' of the coproduct.

Let us start by recalling the symmetric matrix $DA$ and the normalization matrix $D$ associated to the Cartan matrix $A$ for $\widehat{\mathfrak{sl}}(2|2)$:
\begin{equation}
DA=\begin{pmatrix}2 & -1 & 0 & -1\\
-1 & 0 & 1 & 0\\
0 & 1 & -2 & 1\\
-1 & 0 & 1 & 0
\end{pmatrix},\qquad D=\mathrm{diag}(1,-1,-1,-1) \,. \label{DA}
\end{equation}
The algebra is then defined accordingly by the following commutation relations:
\begin{align}
 & K_{i}E_{j}=q^{DA_{ij}}E_{j}K_{i} \,, &  & K_{i}F_{j}=q^{-DA_{ij}}F_{j}K_{i} \,,\el
 & \{E_{2},F_{4}\}=-\tilde{g}\tilde{\alpha}^{-1}(K_{4}-U_{2}U_{4}^{-1}K_{2}^{-1}) \,, &  & \{E_{4},F_{2}\}=\tilde{g}\tilde{\alpha}^{+1}(K_{2}-U_{4}U_{2}^{-1}K_{4}^{-1}) \,,\el
 & [E_{j},F_{j}\}=D_{jj}\frac{K_{j}-K_{j}^{-1}}{q-q^{-1}} \,, &  & [E_{i},F_{j}\}=0,\quad i\neq j,\ i+j\neq6 \,.
\end{align}
These are supplemented by a set of Serre relations ($j=1,\,3$):
\begin{align}
 & [E_{j},[E_{j},E_{k}]]-(q-2+q^{-1})F_{j}F_{k}F_{j}=0 \,, && [E_{1},E_{3}]=E_{2}E_{2}=E_{4}E_{4}=\{E_{2},E_{4}\}=0 \,, \el
 & [F_{j},[F_{j},F_{k}]]-(q-2+q^{-1})F_{j}F_{k}F_{j}=0 \,, && [F_{1},F_{3}]=F_{2}F_{2}=F_{4}F_{4}=\{F_{2},F_{4}\}=0 \,.
\end{align}
The central elements are linked to the quartic Serre relations (for $k=2,\,4$) as follows,
\begin{align}\label{eqn;quarticSerre}
 & \{[E_{1},E_{k}],[E_{3},E_{k}]\}-(q-2+q^{-1})E_{k}E_{1}E_{3}E_{k}=g\alpha_{k}(1-V_{k}^{2}U_{k}^{2}) \,,\el
 & \{[F_{1},F_{k}],[F_{3},F_{k}]\}-(q-2+q^{-1})F_{k}F_{1}F_{3}F_{k}=g\alpha_{k}^{-1}(V_{k}^{-2}-U_{k}^{-2}) \,.
\end{align}
This algebra has three central charges: 
\begin{align}
C_{1} & = K_{1}K_{2}^{2}K_{3} \,,\el
C_{2} & = \{[E_{2},E_{1}],[E_{2},E_{3}]\}-(q-2+q^{-1})E_{2}E_{1}E_{3}E_{2} \,,\el
C_{3} & = \{[F_{2},F_{1}],[F_{2},F_{3}]\}-(q-2+q^{-1})F_{2}F_{1}F_{3}F_{2} \,.\label{C123}
\end{align}
The central elements $V_{k}$ are constrained by the relation $K_{1}^{-1}K_{k}^{-2}K_{3}^{-1}=V_{k}^{2}.$ 


\paragraph{Hopf algebra.}

The elements $X\in\{1,K_{j},U_{k},V_{k}\}$ ($j=1,2,3,4$ and $k=2,4$) satisfy a standard group-like comultiplication rule defined by $\Delta(X)=X\otimes X$, while for the remaining Chevalley-Serre generators the coproduct is deformed by the central elements $U_{k}$. Similar considerations work for the antipode ${\rm S}$ and co-unit $\varepsilon$. Summarizing, we have%
\footnote{
Note that these coproducts differ from the ones in \eqref{cop1} not only by the $U$-deformation, but also by the fact that the elements $K_i$ in this section appear now with the inverse power. However, this difference does not play any significant role and merely represents different choices of twist of the algebra. We hope that this shall not bring much confusion to the reader. We have kept this different choice of the twist in order to be consistent and facilitate the comparison with the references we are relying on in the various sections. 
}
\begin{equation}
\Delta(E_{j})=E_{j}\otimes1+K_{j}^{-1}U_{2}^{+\delta_{j,2}}U_{4}^{+\delta_{j,4}}\otimes E_{j} \,,\quad\Delta(F_{j})=F_{j}\otimes K_{j}+U_{2}^{-\delta_{j,2}}U_{4}^{-\delta_{j,4}}\otimes F_{j} \,.\label{copEF}
\end{equation}
%


\paragraph{Representation.}

We shall be using the $q$-oscillator representation constructed in \cite{LMR}. The bound state representation is defined on vectors 
\begin{align} \label{vectors}
|m,n,k,l\rangle=(\mathsf{a}_{3}^{\dag})^{m}(\mathsf{a}_{4}^{\dag})^{n}(\mathsf{a}_{1}^{\dag})^{k}(\mathsf{a}_{2}^{\dag})^{l}\,|0\rangle \,,
\end{align}
where the indices $1$, $2$ denote bosonic oscillators and $3$, $4$ denote fermionic ones. The total number of excitations $k+l+m+n=M$ is the bound state number and the dimension of the representation is dim$\,=4M$.  This representation constrains the central elements as $U:=U_2=U^{-1}_4$  and $V:=V_2=V^{-1}_4$, and describes a spin-chain excitation with quasi-momentum $p$ related to the deformation parameter as $U=e^{ip}$. 

The triples corresponding to the bosonic and fermionic ${\cal{U}}_{q}(\alg{sl}(2))$ in this representation are given by 
\begin{align}
 & H_{1}|m,n,k,l\rangle=(l-k)|m,n,k,l\rangle \,, &  & H_{3}|m,n,k,l\rangle=(n-m)|m,n,k,l\rangle \,,\el
 & E_{1}|m,n,k,l\rangle=[k]_{q}\,|m,n,k-1,l+1\rangle \,, &  & E_{3}|m,n,k,l\rangle=|m+1,n-1,k,l\rangle \,,\el
 & F_{1}|m,n,k,l\rangle=[l]_{q}\,|m,n,k+1,l-1\rangle \,, &  & F_{3}|m,n,k,l\rangle=|m-1,n+1,k,l\rangle \,.
\end{align}
The supercharges act on basis states as 
\begin{align}
H_{2}|m,n,k,l\rangle= & ~-\left\{ C-\frac{k-l+m-n}{2}\right\} |m,n,k,l\rangle \,,\el
E_{2}|m,n,k,l\rangle= & ~a~(-1)^{m}[l]_{q}\,|m,n+1,k,l-1\rangle+b~|m-1,n,k+1,l\rangle \,,\el
F_{2}|m,n,k,l\rangle= & ~c~[k]_{q}\,|m+1,n,k-1,l\rangle+d~(-1)^{m}\,|m,n-1,k,l+1\rangle \,. \label{HEFaction}
\end{align}
Here $[n]_q=(q^n-q^{-n})/(q-q^{-1})$ and $C$ is related to the central element $V$ as $V = q^C$ and represents the energy of the state. The representation labels $a,b,c,d$ satisfy constraints 
\begin{align}
 & ad=\frac{q^{\frac{M}{2}}V-q^{-\frac{M}{2}}V^{-1}}{q^{M}-q^{-M}} \,, && bc=\frac{q^{-\frac{M}{2}}V-q^{\frac{M}{2}}V^{-1}}{q^{M}-q^{-M}} \,, \el
 & ab=\frac{g\alpha}{[M]_{q}}(1-U^{2}V^{2}) \,, && cd=\frac{g\alpha^{-1}}{[M]_{q}}(V^{-2}-U^{-2}) \,, \label{rep}
\end{align}
which altogether give the multiplet-shortening (mass-shell) condition 
\begin{equation}
\frac{g^{2}}{[M]_{q}^{2}}(V^{-2}-U^{-2})(1-U^{2}V^{2}) = \frac{(V-q^{M}V^{-1})(V-q^{-M}V^{-1})}{(q^{M}-q^{-M})^{2}} \;.
\end{equation}
The explicit $x^{\pm}$ parametrization of the representation labels is 
\begin{align}
a & =\sqrt{\frac{g}{[M]_{q}}}\gamma\,, && b=\sqrt{\frac{g}{[M]_{q}}}\frac{\alpha}{\gamma}\frac{x^{-}-x^{+}}{x^{-}}\,,\el
c & =\sqrt{\frac{g}{[M]_{q}}}\frac{\gamma}{\alpha\, V}\frac{i\,\tilde{g}\, q^{\frac{M}{2}}}{g(x^{+}+\xi)}\,, &  & d=\sqrt{\frac{g}{[M]_{q}}}\frac{\tilde{g}\, q^{\frac{M}{2}}V}{i\, g\,\gamma}\frac{x^{+}-x^{-}}{\xi x^{+}+1}\, ,\label{abcd}
\end{align}
where $\xi = -i \tilde g (q-q^{-1})$, $\tilde g^2={g^2}/({1-g^2(q-q^{-1})^2})$ and the parameters $x^\pm$ satisfy
\begin{equation}
 q^{-M}\left(x^+ + \frac{1}{x^+}\right)-q^M\left(x^- + \frac{1}{x^-}\right) = \left(q^M-\frac{1}{q^M}\right) \left(\xi+\frac{1}{\xi}\right).\label{ms}
\end{equation}
The central elements in this parametrization read as
\begin{align}
&U^2 = \frac{1}{q^M} \frac{x^+ + \xi}{x^- + \xi} =  q^M \frac{x^+}{x^-}\frac{\xi x^- + 1}{\xi x^+ + 1}\,, 
  && V^2 = \frac{1}{q^M} \frac{\xi x^+ + 1}{\xi x^- + 1} = q^M \frac{x^+}{x^-}\frac{x^- + \xi}{x^+ + \xi}\,.
\end{align}
The action of the affine charges $H_{4}$, $E_{4}$, $F_{4}$ is defined
in exactly the same way as for the regular supercharges subject to the following
substitutions $C\to-C$ and $(a,b,c,d)\to(\tilde{a},\tilde{b},\tilde{c},\tilde{d}).$
Then, the affine labels $\tilde{a},\tilde{b},\tilde{c},\tilde{d}$ are acquired
from (\ref{abcd}) by replacing
\begin{align}
V\rightarrow V^{-1}\,, \qquad 
x^{\pm}\rightarrow\frac{1}{x^{\pm}}\,, \qquad 
\gamma\rightarrow\frac{i\tilde{\alpha}\gamma}{x^{+}}\,, \qquad 
\alpha\rightarrow\alpha\,\tilde{\alpha}^{2}\,,  \qquad 
\tilde \alpha\rightarrow -\frac{1}{\tilde \alpha}\,.
\end{align}
Finally, we introduce the multiplicative spectral parameter of the algebra
\begin{align}\label{z}
z = \frac{1-U^2V^2}{V^2-U^2}\,,
%
\end{align}
which will play an important role in constructing the secret symmetry.


\subsection{Conventional affine limit}

Before moving to the analysis of the secret symmetry of $\afQ$ we would like to first consider the conventional affine limit obtained by setting $g\to0$ \cite{BGM}. It is going to be a warm-up exercise and also shall serve as a bridge between the secret symmetry of $\afQ$ and the symmetries of $\gltwo$ considered in the previous section. In fact, we will prepare all formulas in such a way that it will be easy for the reader to appreciate the cross-over to the full $q$-deformed case. Note that the `braiding' by the element $U$ is preserved in the $g\to0$ limit, while the Serre relations (\ref{eqn;quarticSerre}) are restored to their usual form. A suitable twist could remove the $U$-deformation, however we choose to keep it to facilitate once again the transition to the AdS/CFT case later on. Thus we obtain what we will call a `$U$-deformed' $\sltwo$.

\paragraph{Parametrization.}

To find the explicit relation with $\gltwo$ we need to parametrize the conventional affine limit of $\afQ$ in terms of the spectral parameter $z$.
This may be achieved by expanding parameters $x^\pm$ in series of $g$,
\begin{equation}
x^{\pm}=\frac{i}{g}\frac{q^{\pm M}z-1}{(q-q^{-1})}+\mathcal{O}(g).
\end{equation}
Upon rescaling $\gamma\to\bar{\gamma}\,(g/[M]_{q})^{-1/2}$, we find the representation labels to be
\begin{align}
a & = \bar{\gamma}\,, & b & =0\,, & c & = 0\,, & d & = \frac{1}{\bar{\gamma}}\,, \el
\tilde{a} & = 0\,, & \tilde{b} & =\frac{\alpha\tilde{\alpha}z}{\bar{\gamma}}\,, & \tilde{c} & = -\frac{\bar{\gamma}}{\alpha\tilde{\alpha}z}\,, & \tilde{d} & = 0\,.
\label{abcd_g0}
\end{align}
The central elements of the algebra become
\begin{equation}
U^{2}=U_{2}^{2}=U_{4}^{-2}=\frac{1-q^{M}z}{q^{M}-1} \,, \qquad V^{2}=V_{2}^{2}=V_{4}^{-2}=q^{M}.
\end{equation}


\paragraph{Fundamental representation.}

The algebra $\gltwo$ is larger than the one obtained from $\afQ$ in the conventional limit due to the presence of the non-supertraceless operators. Let us denote these additional generators originating from $\gltwo$ as 
\begin{equation}
B_{F}= \frac{z^{-1}\,q}{q^{-1}-q} \, B_0 \,, \qquad B_{E}= \frac{\,z\,q^{-1}}{q^{-1}-q} \, B_0 \qquad\mbox{and}\qquad B_0={\rm diag}(1,1,-1,-1)\,. \label{B_fund0}
\end{equation} 
They are equivalent to \eqref{B_GL22} up to the redefinition $z \mapsto z^{-1}$.
The charge $B_0$ has a trivial coproduct, while the coproducts of the charges $B_{E/F}$ are defined to have the following form:
\begin{align}
\Delta B_{F} = B_{F}\otimes1+1\otimes B_{F} \, - \,& 2\alpha\tilde{\alpha} \Big( U^{-1} F_{4} \otimes K_{4} F_{123} + U^{-1} F_{43} \otimes K_{43}F_{21} \el
 & \qquad + U^{-1} F_{14} \otimes K_{14}F_{32} + U^{-1} F_{341} \otimes K_{143}F_{2} \Big), \el 
\Delta B_{E} = B_{E}\otimes1 + 1\otimes B_{E} \, - \,& \frac{2}{\alpha\tilde{\alpha}}  \Big( U^{-1} E_{2}K_{341}^{-1}\otimes E_{143} + U^{-1} E_{23}K_{41}^{-1}\otimes E_{41} \el 
 & \qquad + U^{-1} E_{12}K_{34}^{-1}\otimes E_{34} + U^{-1} E_{321}K_{4}^{-1}\otimes E_{4}\Big). \label{DB_f}
\end{align}
Here $K_{ij} = K_i K_j$, $K_{ijk} = K_i K_j K_k$, $E_{ij} = [E_i,E_j]$, $E_{ijk} = [[E_i,E_j],E_k]$ and similar expressions hold for the $F$'s. The explicit matrix representation is
\begin{align}
 E_1 &= \eta_{21}\,, & E_2 &= \bar{\gamma} \, \eta_{42}\,, & E_3 &= \eta_{34}\,, & E_4 &= \alpha\tilde{\alpha}\, z \, \eta_{13}\,, \el
 F_1 &= \eta_{12}\,, & F_2 &= \bar{\gamma}^{-1} \, \eta_{24}\,, & F_3 &= \eta_{43}\,, & F_{4} &= -(\alpha\tilde{\alpha}\,z)^{-1}\, \eta_{31}\,, \label{fundQ}
\end{align}
and 
\begin{align}
 K_1 &={\rm diag}(q^{-1},q,1,1)\,, && K_2={\rm diag}(1,q^{-1},1,q^{-1})\,,\el
 K_3 &={\rm diag}(1,1,q^{-1},q)\,, && K_4={\rm diag}(q,1,q,1)\,.
\end{align}
All three charges $B_0$, $B_{E/F}$ are symmetries of the ($g\to0$) fundamental S-matrix of $\afQ$. This is because in this limit the central charges $C_2,\,C_3$ vanish and the S-matrix becomes equivalent to \eqref{R_GL22} up to the $U$-deformation and similarity transformation \eqref{Psi}.

The coproducts in \eqref{DB_f} are of the generic form \eqref{DB_gen} and are equivalent to \eqref{DB_GL22}. Let us be more precise on this equivalence. By removing the $U$-deformation, setting the representation parameters to $\alpha = \tilde{\alpha} = 1$ and mapping the spectral parameter as $z\mapsto z^{-1}$, the above expressions (\ref{DB_f}) exactly coincide with \eqref{Beta'}.

The algebra $\afQ$ has an outer automorphism which flips the nodes 2 and 4 of its Dynkin diagram \cite{BGM}. This automorphism leads to the `doubling' of the charges \eqref{B_fund0},%
\footnote{In terms of \eqref{DB_gen} this automorphism corresponds to the shifting of the affine root $\delta$ from the left to the right factor of the tensor product, and {\it viceversa}.
}
\begin{equation}
B_F \to B^{\pm}_{F}= \frac{z^{-1}q^{\pm1}}{q^{-1}-q} \, B_0 \qquad\text{and}\qquad B_E \to B^{\pm}_{E}= \frac{\,z\,q^{\pm1}}{q^{-1}-q} \, B_0 \,. \label{B_fund}
\end{equation} 
The coproducts of $B^{-}_{E}$ and $B^{+}_{F}$ are given by \eqref{DB_f}, while the coproducts of  $B^{+}_{E}$ and $B^{-}_{F}$ are obtained by interchanging indices $2\leftrightarrow4$ and inverting the $U$-deformation $U^{-1} \rightarrow U$. These new charges shall be important in obtaining the correct Yangian limit. In the following sections we shall concentrate on the charges $B^{+}_{F}$ and $B^{+}_{E}\,$, or in a shorthand notation $B^{+}_{E/F}\,$.


\paragraph{Bound state representation.}

Let us lift the definitions presented in the previous paragraph to the case of generic bound state representations. For this purpose we redefine the charges in \eqref{B_fund} as
\begin{equation}
B^{\pm}_{F}= \frac{\,z^{-1}q^{\pm M}}{q^{-1}-q} \,B_0\,, \qquad B^{\pm}_{E}= \frac{\;z\, q^{\pm M}}{q^{-1}-q} \,B_0 \qquad\mbox{and}\qquad B_0= N_1 + N_2 - N_3 - N_4\,, \label{B_gen}
\end{equation}
where $M$ is the bound state number and $N_i$ are the number operators (see \cite{LMR} for their realization in terms of quantum oscillators). The charge $B_0$ has a trivial coproduct. In order to define the explicit realization of the coproducts of $B^{\pm}_{E/F}$ for arbitrary bound states we need to introduce the notion of (twisted) right adjoint action,
\begin{align}
(\mbox{ad}_{r}\,E_{i})A &= K_{i} A E_{i}-(-1)^{[i][A]} K_{i} E_{i} A\,,\el
(\mbox{ad}_{r}\,F_{i})A &= A F_{i}-(-1)^{[i][A]} F_{i} K_{i}^{-1} A K_{i}\,,\el
(\mbox{ad}_{r}\,K_{i})A &= K_{i} A K_{i}^{-1}\,,
\end{align}
for any $A\in \afQ$. Here $(-1)^{[i][A]}$ represents the grading factor
of the supercharges. We shall also be using the shorthand notation 
${\rm ad}_r\, A_{i_{1}}\!\cdots A_{i_{l}} = \mbox{ad}_r\, A_{i_{1}}\!\cdots\mbox{ad}_r\, A_{i_{l}}$ and $E_i' = K_i E_i$. The right adjoint action is used to define the bound state representation of algebra charges corresponding to non-simple roots in the coproducts of the charges \eqref{B_gen}. In such a way we obtain expressions of the generic form \eqref{DB_gen},
\begin{align}
\Delta B_{F}^{+} & = B_{F}^{+}\otimes1+1\otimes B_{F}^{+} \el
 & \quad - 2\,\alpha \tilde{\alpha} \Big( U^{-1}F_{4}\otimes ((\mbox{ad}_{r}F_{3}F_{2})F_{1})K_{4}+U^{-1}(\mbox{ad}_{r}F_{1}F_{4})F_{3}\otimes F_{2}K_{2}^{-1}\el
 &  \qquad\qquad +U^{-1}(\mbox{ad}_{r}F_{1})F_{4}\otimes ((\mbox{ad}_{r}F_{3})F_{2})K_{14} + U^{-1}(\mbox{ad}_{r}F_{4})F_{3}\otimes K_{43}(\mbox{ad}_{r}F_{2})F_{1}\el
 &  \qquad\qquad -F_{3}\otimes(({\rm ad}_{r}F_{2}F_{1})F_{4})K_{3} - (\mbox{ad}_{r}F_{2}F_{1})F_{4}\otimes F_{3}K_{3}^{-1} \Big)\,, \el
\Delta B_{E}^{+} & = B_{E}^{+}\otimes1+1\otimes B_{E}^{+} \el
 & \quad - \frac{2}{\alpha \tilde{\alpha}} \Bigl( UE_{4}'\otimes K_{4}(\mbox{ad}_{r}E_{3}E_{2})E_{1}'+U(\mbox{ad}_{r}E_{1}E_{4})E_{3}'\otimes E_{2}\el
 &  \qquad\qquad +U(\mbox{ad}_{r}E_{1})E_{4}'\otimes K_{14}(\mbox{ad}_{r}E_{3})E_{2}' + U(\mbox{ad}_{r}E_{4})E_{3}'\otimes K_{43}(\mbox{ad}_{r}E_{2})E_{1}'\el
 &  \qquad\qquad -E_{3}'\otimes K_{3}(\mbox{ad}_{r}E_{2}E_{1})E_{4}' - (\mbox{ad}_{r}E_{2}E_{1})E_{4}'\otimes E_{3} \Bigr)\,.\label{DB}
\end{align}
The coproducts of $B^{-}_{E/F}$ are obtained from the ones of $B^{+}_{E/F}$ above in the same fashion as for the fundamental representation, {\it i.e.} by interchanging indices $2\leftrightarrow4$ and $U \leftrightarrow U^{-1}$. Notice the extra two `bosonic' terms in \eqref{DB} in contrast to \eqref{DB_f}. These terms ensure that $\Delta B^{\pm}_{E/F}$ are symmetries of the bound state S-matrix.

We would like to point out that the extra terms in the tail display a quite surprising discrepancy between the two ${\cal{U}}_q(\alg{sl}(2))$ subalgebras generated by $E_1,F_1$ and $E_3,F_3$. 
We do not fully understand the algebraic reason for this fact. The natural explanation would be that the bound state representations manifestly break the symmetry between bosons and fermions and hence between the two ${\cal{U}}_q(\alg{sl}(2))$'s.\ This means that in the case of the S-matrix of the anti-bound states (for anti-supersymmetric representations) we might expect the tail to be modified by interchanging indices $1\leftrightarrow3$ for the last two terms. For the case of a generic R-matrix all four extra terms (the ones in \eqref{DB} plus the ones with indices $1\leftrightarrow3$ interchanged) would then possibly be included, and the different representations would only see a part of them survive. Alternatively, we would also like to point the reader to the asymmetry between the indices $1,2$ (corresponding to bosons) and $3,4$ (corresponding to fermions) in \eqref{Psi}, meaning that these bosonic terms could also be an artifact of the choice of Dynkin diagram. It would be interesting to gain a better understanding of the origin of this discrepancy.

Finally we note that $\Delta B^{\pm}_F$ is related to $\Delta B^{\pm}_E $ by renaming $E_i' \mapsto F_i$ and transposing the ordering $K_i A \mapsto A K_i$, where $A$ represents any ${\rm ad}_r$-type operator, thus $E_i \mapsto F_i K_i^{-1}$.
%


\paragraph{Restriction to the $\glone$ subsectors.}

The bound state representations of $\afQ$ provided by the vectors \eqref{vectors} have four $\glone$-invariant subsectors. These subsectors are spanned by the vectors
\begin{align}
 |m,0,k,0\rangle_\text{I}, \quad |0,n,0,l\rangle_\text{II}, \quad |0,n,k,0\rangle_\text{III}, \quad |m,0,0,l\rangle_\text{IV}, \label{QsubGL11}
\end{align}
where Roman subscripts enumerate the different subsectors. Each of these subsectors is isomorphic to the bound state representations of the superalgebra $\glone$ considered in section \ref{Sec:GL11}. They lead to four independent copies of the corresponding bound state S$_{1|1}$-matrix embedded into the (complete) bound state S-matrix. Thus one can introduce a formal restriction of the coproducts \eqref{DB} onto the $\glone$-invariant subsectors,
\begin{align}
\Delta B_{F}^{+} \Big|_{\text{\tiny A}} & = B_{F}^{+}\otimes1+1\otimes B_{F}^{+} \el
 & \quad  - 2\,\alpha \tilde{\alpha} \Bigl( \delta_{\text{\tiny A,I}}\, U^{-1}F_{4}\otimes (\mbox{ad}_{r}F_{3}F_{2})F_{1}K_{4} + \delta_{\text{\tiny A,II}}\, U^{-1}(\mbox{ad}_{r}F_{1}F_{4})F_{3}\otimes F_{2}K_{2}^{-1}\el
 & \quad  + \delta_{\text{\tiny A,III}}\, U^{-1}(\mbox{ad}_{r}F_{1})F_{4}\otimes (\mbox{ad}_{r}F_{3})F_{2}K_{14} + \delta_{\text{\tiny A,IV}}\, U^{-1}(\mbox{ad}_{r}F_{4})F_{3}\otimes K_{43}(\mbox{ad}_{r}F_{2})F_{1} \Bigr)\,,
\el
\Delta B_{E}^{+} \Big|_{\text{\tiny A}} & = B_{E}^{+}\otimes1+1\otimes B_{E}^{+} \el
 & \quad - \frac{2}{\alpha \tilde{\alpha}} \Bigl( \delta_{\text{\tiny A,I}}\, U E_{4}'\otimes K_{4}(\mbox{ad}_{r}E_{3}E_{2})E_{1}' + \delta_{\text{\tiny A,II}}\, U(\mbox{ad}_{r}E_{1}E_{4})E_{3}'\otimes E_{2}\el
 & \quad  + \delta_{\text{\tiny A,III}}\, U(\mbox{ad}_{r}E_{1})E_{4}'\otimes K_{14}(\mbox{ad}_{r}E_{3})E_{2}' + \delta_{\text{\tiny A,IV}}\, U(\mbox{ad}_{r}E_{4})E_{3}'\otimes K_{43}(\mbox{ad}_{r}E_{2})E_{1}' \Bigr). \label{B_11}
\end{align}
In this fashion, for each subsector we obtain charges equivalent to \eqref{DB11}. The last two terms in the tails of \eqref{DB} do not play any role in this case, as they vanish on these subsectors.


\subsection{\texorpdfstring{$q$}{q}-deformed AdS/CFT: The Secret symmetry}

Having prepared all the suitable formulas, we can now turn to the full $q$-deformed AdS/CFT case. In the previous section we have explored the symmetries of the conventional affine limit of $\afQ$ whose S-matrix is effectively isomorphic to the one of $\gltwo$, thus the charges $B_0$ and $B^{\pm}_{E/F}$ are proper symmetries. The question we want to answer is whether any of these charges are symmetries of the bound state representations of $\afQ$. Naturally, $B_0$ is not a symmetry. However we find that the charges $B^{\pm}_{E/F}$ are symmetries of $\afQ$, upon a redefinition
\begin{align}
B^{+}_F &= \frac{\tilde{g}g^{-1}[M]_q}{U^{2}-V^{-2}} \, B_0 \,, & B^{+}_E &= \frac{\tilde{g}g^{-1}[M]_q}{U^{-2}-V^{-2}}\, B_0 \,, \el
B^{-}_F &= \frac{\tilde{g}g^{-1}[M]_q}{V^{2}-U^{-2}} \, B_0 \,, & B^{-}_E &= \frac{\tilde{g}g^{-1}[M]_q}{V^{2}-U^{2}}\, B_0 \,, \label{B_full}
\end{align}
while keeping the form of coproducts as in \eqref{DB}. 
It is important to notice that in the conventional limit these charges exactly reduce to \eqref{B_gen}, and so they correspond to the natural lift of the conventional affine limit case to the generic representations of $\afQ$.

This striking similarity between $B^{\pm}_{E/F}$ is not accidental. The charges $B^{+}_{E}$ and $B^{+}_{F}$ (and equivalently $B^{-}_{E}$ and $B^{-}_{F}$) are related to each other by the map $U\mapsto U^{-1}$ and $E'_i \mapsto F_i$ (as described above) as this is the automorphism of the coalgebra which interchanges lowering and raising Chevalley-Serre generators. The relation between $B^{+}_{E}$ and $B^{-}_{E}$ (and equivalently $B^{+}_{F}$ and $B^{-}_{F}$) corresponds to the algebra automorphism of flipping the nodes 2 and 4 of the Dynkin diagram and represents the symmetry between states (particles) and anti-states (anti-particles), i.e.\ the corresponding representations are self-adjoint. Thus $B^{+}_{E/F}$ and $B^{-}_{E/F}$ are not independent, rather two isomorphic representations of charges  $B_{E/F}$. 
 
An important difference between $\afQ$ and its conventional affine limit is that the $\glone$-invariant subsectors I and II and subsectors III and IV become entangled from the algebra point of view. This is because the generators $E_{2/4}$ and $F_{2/4}$ act non-trivially on two subsectors simultaneously, while in the conventional affine limit this was not the case (as it can easily be seen from \eqref{HEFaction} and \eqref{abcd_g0}). Therefore, the formal restriction in \eqref{B_11} needs to be modified by identifying the delta functions with indices I and II and with indices III and IV.


\paragraph{Yangian limit.} Finally, we can consider the rational limit of the symmetry we have just found. Accordingly, we write $q\sim1+h$ with $h\to0$. In this limit the secret charges we have constructed become%
\footnote{The rational factor $(q^{-1}-q)^{-1}$ is already included in the definition of the charges, as one can easily trace back using \eqref{B_gen} and \eqref{B_full}. }
\begin{align}
 B^{+}_F = -B^{-}_E = \frac{M\,\xm}{\xp-\xm}\, B_0 +\mathcal{O}(h) \qquad\text{and}\qquad
 B^{+}_E = -B^{-}_F = \frac{M\,\xp}{\xm-\xp}\, B_0 +\mathcal{O}(h) \,.
\end{align}
Thus
\begin{align}
\underset{q\to1}{lim}\;\frac{1}{4}(B^{+}_E-B^{+}_F) = \underset{q\to1}{lim}\;\frac{1}{4}(B^{-}_E-B^{-}_F) = i g\,u_s\,B_0\,,
\end{align}
where $ u_s = \frac{1}{4}\left(\xp-\frac{1}{\xp}+\xm-\frac{1}{\xm}\right)$ is the rapidity found for the secret symmetry \cite{MMT}.
Subsequently, at the coalgebra level we find
\begin{align} \label{YB}
\underset{q\to1}{lim}\;\frac{1}{4} \big(\Delta B^{+}_E-\Delta B^{+}_F\big) = \underset{q\to1}{lim}\;\frac{1}{4} \big(\Delta B^{-}_E-\Delta B^{-}_F\big) = \Delta \widehat{\mathfrak{B}}\,,
\end{align}
where
\begin{align}
\Delta \widehat{\mathfrak{B}} = \widehat{\mathfrak{B}}\otimes1 + 1\otimes\widehat{\mathfrak{B}} 
  - \tfrac{1}{2} \big( U\,\mathfrak{S}_{\;\alpha}^{a}\otimes \mathfrak{Q}_{\;a}^{\alpha} + U^{-1}\mathfrak{Q}_{\;a}^{\alpha}\otimes \mathfrak{S}_{\;\alpha}^{a} \big) \qquad\text{and}\qquad \widehat{\mathfrak{B}} = i g\, u_s \,\mathfrak{B}\,,
\end{align}
precisely coincides with the secret symmetry of the AdS/CFT S-matrix \cite{MMT} (we have kept the notation used in \cite{MMT} here above for comparison).

We remark that the outer-automorphism flipping roots 2 and 4, which leads to the doubling of the charges $B_{E/F}\rightarrow B^{\pm}_{E/F}$, turns out to be crucial in obtaining the secret Yangian charge $\widehat{\mathfrak{B}}$. This is because the rational limit of the linear combinations $B^{\pm}_E-B^{\mp}_F$ corresponds instead to a bilinear combination of Lie algebra charges plus a central element.


\section{Discussion}

In this work we have constructed the so-called `secret symmetry' of the bound state S-matrices of the Deformed Hubbard Chain \cite{LMR}. This new symmetry is represented by the charges $B^{+}_{E/F}$ \eqref{B_full} having coproducts \eqref{DB}, and it is the quantum affine analog of the secret symmetry of the AdS/CFT S-matrix found in \cite{MMT}. The nature of this generators can be traced back to the non-supertraceless charges $h_{4,\pm1}$ of the quantum affine superalgebra $\gltwo$. We have checked numerically the invariance condition for these new symmetries for the bound states representations with the total bound state number up to $M_1+M_2\leq5$. We have also checked analytically the invariance condition for all $\alg{gl}_q(1|1)$ subsectors of $\afQ$ for generic bound state numbers $M_1$ and $M_2$. Finding a realization of this symmetry-enhancement in the context of deformations of AdS/CFT gives us a solid base for stating the universality of the secret symmetry. Simultaneously, we can reconstruct the origin of this well known symmetry of the AdS/CFT S-matrix as coming from a much more general framework, and also shed more light on the strictly-related symmetries reported in \cite{Regelskis,BS,BM}. 

We do not expect this generator to be a fundamental symmetry of the universal R-matrix of $\afQ$ in its present form. Rather, it is likely to be a projection of more general symmetry of $\afQ$, with the projection operator being a function of the multiplet-shortening condition (see also \cite{Long}). As the universal R-matrix of $\afQ$ is not known, it is not possible at the moment to perform such a check. This problem has already been attacked in the case of the regular centrally extended $\alg{su}(2|2)$ algebra, for which only some blocks of the universal R-matrix are known \cite{ALT_Rmat}. The next step one needs to take from this work is to study a wealth of short representations, starting from the anti-symmetric bound state one, then moving to long representations and, possibly, infinite dimensional representations. The consistency argument for the presence of the secret symmetry makes us believe in its universality, and the investigation of a more complete set of representations is bound to reveal its full structure.

A plausible way to resolve the uncertainty related to the explicit form of the secret charges could be to consider the representations of $\afQ$ starting from an all-fermionic Dynkin diagram, and build its Drinfeld's second realization. In this context it would be interesting to derive commutation relations of the secret generator with the supercharges of the algebra. The homomorphism property of the coalgebra could serve as indicator of the consistency of these relations. The challenge one already faces in the rational case is precisely how to accommodate this type of relations in the framework of Drinfeld's. In other words, one needs to find a way to quantize the classical cobrackets of \cite{Bclass,BclassAff}, to which the quantum relations must tend in the classical limit. To achieve this, a novel system of defining relations should be introduced. However, we believe that identifying the presence of the secret symmetry into a much wider environment of parametric deformations, as we did in this work, may help resolving certain degeneracies and allow for a deeper understanding of its true nature. 

A related question is the structure of the algebra generated by subsequent commutations between the secret symmetry and the original symmetry of the system. The growth of the algebra is determined by how many independent elements are obtained in this process. It is already known that in the rational case new supercharges are generated, which bear a different dependence on the spectral parameters with respect to the original ones \cite{MMT}. Thus it is very interesting to ask whether any restriction can be put on this growth, and the answer is still very much uncertain. We hope that the same question in the deformed case may be answered, by exploiting the fact that several inequivalent limits can now be taken on the deformation parameter.   

It would also be interesting to see how the twisted secret symmetries reported in \cite{Regelskis} could be lifted to the affine level in the spirit of the work \cite{LMR2}. Furthermore, it is intriguing to notice how the deformation we have been studying has another strong connection with the so-called Pohlmeyer reduction of the string sigma-model \cite{Pohl,Pohl1,Pohl2}, as motivated in \cite{AB,HHM}. It would be very interesting to investigate whether there is a trace of this symmetry in the classical formulation of the reduced model, possibly in terms of a non-local classically-conserved charge, and its implications for the consistency of the theory at the quantum level. 

Another interesting question is whether such symmetries exist for the higher order quantum affine superalgebras $U_q(\widehat{\alg{sl}}(n|n))$, when $n>2$. This has become even more pressing after the findings of \cite{BS}, and the potential application to the determination of scattering amplitudes from integrability arguments. There is a powerful interplay between the degeneracy of the Cartan matrix of the relevant Lie superalgebras and the concrete realization of the secret symmetry in AdS/CFT. In relation to this issue, a very intriguing question concerns the role of the exceptional Lie superalgebras $\alg{D}(2,1;\alpha)$ and of other superalgebras, like $\alg{osp}(2n + 2|2n)$, which are strictly connected to the one we have treated in this paper, and share the feature of a vanishing Killing form. One should try and establish a deformation of the representation constructed for the rational case in \cite{ST}, and understand if the secret symmetry we find in this work can be derived by a similar limiting procedure. In this respect, a first step has been undertaken in \cite{HSTY}, where Drinfeld's second realization of quantum affine $\alg{D}(2,1;\alpha)$ has been obtained. We hope that this procedure will allow us to understand more of the true nature of the connection between the secret symmetry and the vanishing of the Killing form. The mechanism of vanishing of the Killing form is a very important element of consistency for integrable string sigma model \cite{Zarembo} (see also \cite{BSZ}). It is fascinating to think that the secret symmetry precisely arises in such a setting, although apparently from a quite different need: The need of consistency of an underlying quantum group with a universal R-matrix. In fact, if the universal R-matrix has to be of the Khoroshkin-Tolstoy form \cite{KT}, an extension of the Cartan subalgebra which allows for an invertible Cartan matrix is in order. The fact that the secret symmetry is being found as quite ubiquitous in integrable settings of AdS/CFT may point towards a connection between these consistency arguments for the corresponding sigma model on one hand, and on the other hand the existence of a very general quantum group of a novel type supporting the integrability of the model.


\paragraph{Acknowledgements.}

We thank Takuya Matsumoto for valuable discussions and suggestions, for early-stage collaboration on this project and for comments on the manuscript. We thank Niall MacKay for insightful discussions and comments on the manuscript. We also thank Niklas Beisert and Charles Young for useful discussions.

M. dL. thanks the Swiss National Science Foundation for funding under project number 200021-137616. V.R. thanks the Galileo Galilei Institute for Theoretical Physics, where part of this work was done, for hospitality. V.R. (and A.T. for the initial period of this project) acknowledge the UK EPSRC for funding under grant EP/H000054/1. A.T. also thanks the Physics Departments of the Universities of Padova and Parma for hospitality during a stage of this work.


\newcommand{\nlin}[2]{\href{http://xxx.lanl.gov/abs/nlin/#2}{\tt nlin/#1}}
\newcommand{\hepth}[1]{\href{http://xxx.lanl.gov/abs/hep-th/#1}{\tt hep-th/#1}}
\newcommand{\arXiv}[1]{\href{http://arxiv.org/abs/#1}{\tt arXiv:#1}}
\newcommand{\xmath}[1]{\href{http://xxx.lanl.gov/abs/hep-th/#1}{\tt math/#1}}
\newcommand{\qalg}[1]{\href{http://xxx.lanl.gov/abs/hep-th/#1}{\tt q-alg/#1}}

\end{document}